\documentclass[journal = jpccck, manuscript = article]{achemso} 
\setkeys{acs}{usetitle = true}
\usepackage{graphicx}
\usepackage{epsfig}
\usepackage{bm} 
\usepackage[latin1]{inputenc} 
\usepackage{amsmath,amsfonts,amssymb,amsthm,mathrsfs}
\usepackage{natbib} 
\title{
Resonant Lifetime of 
Core-Excited 
Organic Adsorbates 
from 
First Principles}
\author{Guido Fratesi}\email{guido.fratesi@unimi.it}
\affiliation{ETSF, CNISM, Dipartimento di Fisica, Universit\`a degli Studi di Milano, via Celoria 16, I-20133 Milano, Italy}
\altaffiliation{Dipartimento di Scienza dei Materiali, Universit\`a di Milano-Bicocca, via Cozzi 55, I-20125 Milano, Italy}
\author{Carlo Motta\footnote{Current address: School of Physics and CRANN, Trinity College, Dublin 2, Ireland}}
\affiliation{Dipartimento di Scienza dei Materiali, Universit\`a di Milano-Bicocca, via Cozzi 55, I-20125 Milano, Italy}
\author{Mario Italo Trioni}
\affiliation{CNR-National Research Council of Italy, ISTM, Via Golgi 19, I-20133 Milano, Italy}
\author{Gian Paolo Brivio}
\affiliation{Dipartimento di Scienza dei Materiali, Universit\`a di Milano-Bicocca, via Cozzi 55, I-20125 Milano, Italy}
\author{Daniel S\'anchez-Portal}
\affiliation{Centro de F\'{\i}sica de Materiales CSIC-UPV/EHU, 
Paseo Manuel de Lardizabal 5, 20018 San Sebasti\'an, Spain}
\altaffiliation{ 
Donostia International Physics Center (DIPC), Paseo Manuel de Lardizabal 4, 20018 San Sebasti\'an, Spain}
\begin{document}

\begin{abstract}
We investigate by first-principles simulations the resonant 
%%DSP contribution to the 
electron-transfer lifetime from the excited state of an organic adsorbate 
to a semiconductor surface, namely isonicotinic acid on rutile TiO$_2$(110). 
%%DSP Coupling of the initial, single-molecule localized state, to the 
The molecule-substrate interaction is described using density functional theory, while 
the effect of a truly semi-infinite substrate is taken into account 
by Green's function techniques. 
%Our computational scheme is based on widely available methodologies
%for the evaluation of ballistic
%conductance in molecular junctions and nanostructures. 
%%, for which we propose a numerically viable Green's function based on widely available methodologies for the evaluation of ballistic conductance through nano-contacts.
Excitonic effects due to the presence of core-excited atoms in the molecule
are shown to be instrumental to understand the electron-transfer times measured using 
the so-called core-hole-clock technique.
In particular, for the isonicotinic acid on TiO$_2$(110), we find that the charge injection
from the LUMO is quenched since this state lies within the substrate band gap.   
%%DSPInteraction with the core hole, as it results when the molecular orbital is populated by X-ray absorption, is described in a single-particle picture and shown to be of fundamental importance when making reference to experimental results, as molecular states could shift in the substrate band-gap.  
We compute the resonant charge-transfer times from LUMO+1 and LUMO+2,
and systematically investigate the dependence of the elastic lifetimes of these states
on the alignment among adsorbate
 and substrate states. 
\end{abstract}
{{\bf Keywords:} Dye sensitized solar cells; Organic electronics; Charge-transfer dynamics; Resonant spectroscopy; Density functional theory}  %%%JPCC
%%% \maketitle %%%APS
\section{\label{sec:introduction}Introduction}
Charge transfer from an electronically excited adsorbed molecule is a process with wide relevance in several fields such as surface reaction dynamics, photocatalysis, molecular electronics and organic photovoltaics. Regarding the last topic, dye 
sensitized solar cells (DSSC) are an important emergent technology providing new ways 
%%DSP for electricity generation 
%%DSPwhich harvest light energy
to harvest energy from light.\cite{10.1016/S1389-5567(03)00026-1, 10.1021/jp2026847, 10.1021/cm200651e, 10.1038/nphoton.2012.22} They are 
%%DSP
typically 
%%DSP
constituted by a transition metal complex or an organic dye adsorbed on 
%%DSP
an insulating
%%DSP 
substrate, which in most cases is TiO$_2$ or ZnO. Light radiation excites 
%%DSP
%%an electron 
%%from an occupied state (usually the HOMO of the adsorbate) 
an electronic transition within the adsorbate, usually involving the highest occupied molecular orbital (HOMO),
%%to an excited state of the dye which
and the excited electron 
may be transferred to the continuum of states in the conduction band of the substrate. 
%%DSP
The oxidized dye is then restored to its reduced form by a redox pair in solution, and a photocurrent can be generated in the circuit. The 
%%DSP
relative
time scales of these processes are fundamental
%%DSP to allow 
for the cell to work properly. In particular, the first step, i.e., the transfer of the excited electron to the anode, must be faster than charge recombination within the molecule 
%%DSP
so the electron can be injected to the surface and quickly delocalize into the bulk.  In this way, the importance of the recombination channel within the molecule is reduced and the produced electron and hole
can be efficiently separated.
%%DSP

%%DSP first reference to unoccupied states to justify the inclusion of IP at 
%%the same level as 2PPE, we could also add a reference about NEXAFS to 
%% the core-level spectroscopies 
Unoccupied levels and electron charge-transfer dynamics at surfaces can be investigated by a variety of experimental techniques
%%DSP: 
like resonant core-level electron spectroscopy,\cite{10.1103/RevModPhys.74.703, 10.1039/b719546j,10.1016/S0301-0104(99)00305-5} 
%%DSP
two-photon photoemission and other femtosecond pump-probe 
time-resolved (PPTR) laser spectroscopies,\cite{10.1021/jp971581q, 10.1021/jp052078d, 10.1088/1367-2630/14/4/043023} and
%% DSP
%%Do you think that inverse photoemission gives information about
%% electron dynamics?, I believe not, the resolution is typically too low to 
%% get widths. am I wrong?
 inverse photoemission spectroscopy (IPS).\cite{10.1016/S0368-2048(99)00058-4, 10.1016/S0009-2614(99)01384-6, 10.1016/j.carbon.2011.12.055}
In the first method, also named core-hole-clock
%%DSP
spectroscopy, 
a shallow core level of the absorbed molecule is photoexcited by X-ray absorption (XAS) 
%%DSP
into 
an unoccupied bound state 
%%DSP of the interacting system
of the adsorbate.
The photoexcited electron is eventually injected to the substrate, while
%%which remains charge neutral until electron is injection to the substrate 
the decay of the excited core usually proceeds via an Auger decay.
The Auger electron can be emitted before or after the photoexcited electron 
(sometimes referred to as spectator electron)
is transferred to the surface. From the relative intensities of these two Auger decay
channels, the
charge-transfer time of the initially excited electron can be obtained
in units of the Auger decay time. 
%%The Auger emitted electrons are detected and from their energies and  it can 
This technique has the advantage to access the fastest timescales 
%%DSP
(down to the attosecond\cite{10.1038/nature03833}), 
but the core hole produces a significant perturbation to the valence electronic structure.
%%DSP
For this reason, unoccupied states might significantly differ from that of the molecule excited from more extended molecular orbitals (MOs), like the HOMO. For example, 
%%DSP
the lowest unoccupied molecular orbital (LUMO)  of bi-isonicotinic and 
isonicotinic acid on rutile TiO$_2$(110)  
%%DSP
has been found to lie in the substrate gap
%%DSP
in XAS experiments, since it is lowered by the attractive potential due to the 
%%DSP
core-hole localized in the nitrogen atom.\cite{10.1038/nature00952}
Still, electron transfer times of a few femtoseconds
could be measured for the next two higher resonances 
that overlap with the conduction band.\cite{10.1038/nature00952, 10.1063/1.1586692} 
Another indication of the  strong modification of the molecular electronic structure is due to
the fact that, 
%%DSP
even in relatively small molecules like paracyclophanes, charge transfer times measured by core-hole-clock spectroscopy strongly depend on the site of the core excitation.\cite{10.1038/ncomms2083}

Given the complexity of the dye-substrate system, density functional theory (DFT) is most often the method of choice for the theoretical description of its electronic structure, while the nuclei are described classically. Within DFT, different 
approaches were proposed 
%%DSP focusing on 
to study some of the various aspects of the electron injection process.\cite{10.1039/c1ee01906f, 10.1021/jz400046p}
%%DSP
Estimates of adiabatic and non-adiabatic contributions were accomplished by coupled electron/ion dynamics at various temperatures\cite{10.1021/jp014267b, 10.1146/annurev.physchem.58.052306.144054} or calculation of phonon self-energies,\cite{10.1103/PhysRevB.68.195422} while electron-electron contributions have also been found to be dominant in some cases.\cite{10.1103/PhysRevB.80.195419}  Other methods focus instead on the resonant electron transfer which operates for molecular states well within the energy continuum of the 
%%DSPsolid 
substrate bands.\cite{10.1039/c1ee01906f, 10.1021/jp2026847, 10.1103/PhysRevB.85.235132}

%
% A
%
Albeit cluster\cite{10.1063/1.480945, 10.1021/ct050141x, 10.1021/jp072217m} or slab\cite{10.1021/ja902833s, 10.1021/jp2026847, 10.1021/jz400046p} models are commonly adopted, a truly continuum 
%%DSP band 
density of states, without artificial confinement effects, is only reached for a substrate which extends semi-infinitely also in the direction perpendicular to the surface.\cite{10.1088/0953-8984/19/30/305005}
%
% D
%
Several methods have been proposed over the years to deal with calculations
involving semi-infinite substrates. Among them, one of the most powerful
is probably the so-called embedding technique,\cite{10.1088/0022-3719/14/26/015, 10.1103/RevModPhys.71.231, 10.1088/0953-8984/19/30/305005} 
which allows performing calculations of the surface Green's function
where only a few atomic layers closer to the surface are taken into
account explicitly.  
 %%We recall that a semi-infinite substrate could be taken into account by the embedding method %%for the Kohn-Sham (KS) Green's function.
Using this method, the 
elastic lifetimes of adsorbed molecules on metals have been calculated with great accuracy.\cite{10.1103/PhysRevB.81.165444, 10.1016/j.progsurf.2013.03.002}
%%DSP{\bf [MIT]} % Mario, could you say something about difficulties to implement embedding in complex systems/interfaces? Like defining embedding surface, whatever?
Unfortunately, in practice
the implementation of embedding can be quite
cumbersome and, in many cases, it is
restricted to deal with 
particular geometries that allow performing necessary
simplifications. This has probably prevented
a wider application of this method so far.
%
% B
%
Most approaches 
%%DSP
to date 
%%DSP
also neglect the strong interaction of the valence states with the core-hole, 
%%DSP
%%when making references to
even when trying to account for 
experimental 
results obtained by the core-hole-clock method.
%
% C'
%
These two relevant problems 
%%DSP await for a theoretical description 
need a more careful theoretical treatment.

%
% C''
%
In this paper,
%%DSP by using MO 
we would like to contribute 
%%DSP to that understanding.
along these research lines. 
In particular, we 
%%DSP wish to 
calculate the resonant lifetimes of 
%%DSP 
core-excited organic species, 
%%DSP 
focusing 
on isonicotinic acid adsorbed on rutile TiO$_2$(110) as a relevant example, and compare them
with the experimental information from
core-hole-clock spectroscopy at the N K-edge.\cite{10.1063/1.1586692}
%
% E
%
To deal with a semi-infinite substrate
%%DSP
%As a simpler alternative, 
%%DSP
%%we propose here to  adopt
we use here a Green's function procedure similar to that used by one of us to study resonant charge-transfer from simple adsorbates.\cite{10.1103/PhysRevB.76.235406,10.1016/j.progsurf.2007.03.008} 
%%DSP
We show that such methodology can be easily extended to treat more complex adsorbates,
and different types of substrates,
using the plethora of available techniques for the first-principles
description of quantum transport in 
nanostructures.\cite{10.1103/PhysRevB.65.165401,10.1038/nmat1349,10.1103/PhysRevB.73.235419,10.1142/5703}
%
% F
%
%%F%%Furthermore, although our previous calculations\cite{10.1103/PhysRevB.76.235406,10.1016/j.progsurf.2007.03.008}  were restricted to combine information from Kohn-Sham (KS) Hamiltonians obtained from both bulk and slab calculations, our scheme can be implemented using quantum transport codes that can deal with finite bias drops.
%%F%%This allows, at least in principle, carrying out self-consistent calculations of the surface region under the open boundary conditions imposed by the presence of the semi-infinite substrate. 
 %%  on widely available techniques for the description of conductance through nano-contacts.
%%DSP I added a few references myself. 
%%{\bf [CM]} % Carlo, I guess you have good references here easily at hand for other codes and approaches.
%
% G
%
%%DSP We
%In this paper we
We
also deal with the core hole in the adsorbate, and the resulting excitonic effects. 
Given the large degree of localization of the core hole, one can treat the electron-hole interaction statically by relaxing the electronic structure in the presence of a core-to-valence excitation.\cite{10.1103/PhysRevB.58.8097, 10.1063/1.480945, 10.1038/nature00952, 10.1103/PhysRevLett.96.215502}
In this way we 
%%DSP could 
can account for the large relaxations of some of the MOs induced by 
the presence of the localized hole, as well as the associated changes
in the MO
energies
relative to the substrate bands.
% As shown below, for some MOs this can 
%cause radical changes in the coupling to the substrate and, therefore, in the calculated
%resonant lifetimes. The message of our calculations is then clear and twofold.
%On the one hand, the theoretical
%method must be adapted to the type of data that one wants to describe: 
%%GF While the Hamiltonian obtained from standard  ground-state KS calculations 
%%GF can in many cases be a reasonable approximation to describe the dynamics after valence excitations, like those 
%%GF that take place in PPTR experiments, 
%the explicit inclusion of the core-excitation in the calculation can be critical 
%to obtain values that can be compared with core-hole-clock spectroscopy data.
%%account for XAS lifetimes, while standard ground state calculations within the KS equation %%should better refer to TR and IPS data, within the applicability of MO to excited states.
%%DSP
%On the other hand, different types of experiments might provide very different 
%charge transfer times since, in some cases, they look into
%very different spatial distributions and couplings to the substrate of the involved MOs.

%%DSP (is this necessary?)
%%The organization of the paper is as follows. In Section~\ref{sec:method} we shall outline the %%Green function method used in this paper. Section~\ref{sec:results} will present the calculated %%results, which are further discussed in Sec.~\ref{sec:discussion}. Finally, %%Section~\ref{sec:conclusions} will be devoted to present our conclusions.

\section{\label{sec:method}Method}

\begin{figure}
\includegraphics[width=8.5cm]{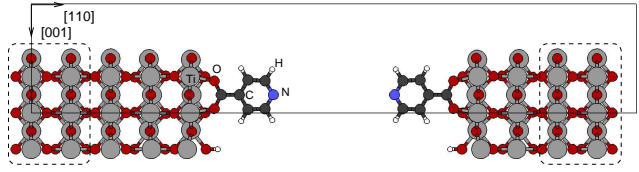}
\caption{\label{fig:geometry}Side view of the model geometry for isonicotinic acid on rutile TiO$_2$(110). The two leftmost and rightmost TiO$_2$ trilayers are coupled to bulk by means of self-energy operators.}\end{figure}

%\subsection{Quantum-transport}
We now describe an {\it ab initio} method to evaluate the resonant lifetime of the MOs of molecules adsorbed at the surface of a semi-infinite crystal. We adopt a Green's function formalism within DFT\cite{10.1103/PhysRevB.76.235406,10.1016/j.progsurf.2007.03.008} 
and setup the problem in analogy to the study of quantum conductance through molecular junctions. Here we make use of the SIESTA/TranSIESTA packages,\cite{10.1103/PhysRevB.65.165401, 10.1088/0953-8984/14/11/302} but we would like to remind that the procedure below is easily realized 
%%DSP by 
using other quantum-transport codes.
%%DSPWe replicate the adsorbate/substrate system as 
We form a fictitious molecular junction
as shown in Fig.~\ref{fig:geometry} and 
%%DSP
assume that the space is divided into three regions: two semi-infinite ``electrodes'' and a central ``contact'' region. The contact region contains two identical copies of the surface under study
and includes a vacuum portion large enough to decouple the left and right adsorbates. 
Notice that, although not strictly necessary,  
it is advisable to choose a symmetric setup like the one depicted in Fig.~\ref{fig:geometry} for reasons outlined below.
The Green's function is computed only in the contact region, upon adding to the Hamiltonian 
the effect of the semi-infinite substrates/electrodes using self-energy operators.
%%DSP This sentence is difficult to understand, so I ahve reomoved it:
 %%which couple the outer space to the boundaries of that region (see the layers marked by the dashed lines).
These are obtained from bulk calculations of the substrate material, as usual 
in transport calculations.\cite{10.1103/PhysRevB.65.165401}
Improving upon our previous calculations,\cite{10.1103/PhysRevB.76.235406,10.1016/j.progsurf.2007.03.008} restricted to combine information from Kohn-Sham (KS) Hamiltonians obtained from both bulk and slab calculations, the current scheme is implemented using quantum transport codes that can deal with finite bias drops. This allows, at least in principle, carrying out self-consistent calculations of the surface region under the open boundary conditions imposed by the presence of the semi-infinite substrate, possibly with the application of external electric fields.
%

%\subsection{Green's function}
%%DSP I present first why we want the Green's function. 
The Green's function is the central quantity in our approach. 
In terms of the Green's function of the surface region we can, for example, compute the 
density of states (DOS) projected onto a wavepacket
$\Phi(\mathbf{r})$ localized at the surface
\begin{equation}\label{eq:DOSPhi}
\rho_{\Phi}(E)=\frac{1}{\pi}\text{Im} \left[ G_{\Phi\Phi}(E) \right]
= \frac{1}{\pi}\text{Im} \left[\int d\mathbf{r} \int d\mathbf{r}' \Phi^*(\mathbf{r}) G(\mathbf{r},\mathbf{r}',E)  \Phi(\mathbf{r}') \right].
\end{equation}
$\rho_{\Phi}(E)$ can be used to treat the dynamics of the population of 
$\Phi(\mathbf{r})$  in several
ways.\cite{10.1103/PhysRevB.76.235406,10.1016/j.progsurf.2007.03.008}  
In particular, if the $\rho_{\Phi}(E)$ is accurately described by a single peak, its
full width at half maximum (FWHM) $\Gamma$ can be used to obtain a good estimation of
the resonant lifetime of the  initial wavepacket $\Phi(\mathbf{r})$ as $\tau=\hbar/\Gamma$.\cite{9780191523878,10.1103/PhysRevB.76.235406}

%%DSP Assuming molecular overlayers with in-plane periodicity, 
Observe that our system is periodic along the in-plane directions. Therefore, 
the Green's function of the full adsorbate/substrate system can be expressed as 

\begin{equation} 
G(\mathbf{r},\mathbf{r}',E)=\frac{1}{N_{\mathbf{k}_\parallel}}
\sum_{\mathbf{k}_\parallel} \sum_{\mu\nu}
G^{\mu\nu}_{\mathbf{k}_\parallel}(E)
\phi_{\mu\mathbf{k}_\parallel}(\mathbf{r})
\phi_{\nu\mathbf{k}_\parallel}^*(\mathbf{r}'),
\end{equation}
where $\mathbf{k}_\parallel$ is the %DSP surface wave-vector 
crystalline momentum parallel to the surface and
\begin{eqnarray} \label{Blochbasis}
\phi_{\mu\mathbf{k}_\parallel}(\mathbf{r})= \sum_{\mathbf{R}} \mathrm{e}^{-i\mathbf{k}_\parallel\cdot(\mathbf{R}+\mathbf{R}_\mu)} \phi_\mu(\mathbf{r}-\mathbf{R}-\mathbf{R}_\mu)
\end{eqnarray}
are Bloch-like basis functions built by summing localized (atom-centered) orbitals $\phi_\mu$ over the Bravais lattice vectors $\mathbf{R}$, with $\mathbf{R}_\mu$ the coordinate 
%DSP of the atom which we associate 
of the center of the $\mu$-th orbital within the unit cell.
The coeffients $G^{\mu\nu}_{\mathbf{k}_\parallel}(E)$ are evaluated from the equation 
\begin{equation}
\sum_{\lambda}G^{\mu\lambda}_{\mathbf{k}_\parallel}(E)[H_{\lambda\nu\mathbf{k}_\parallel}-ES_{\lambda\nu\mathbf{k}_\parallel}]=\delta^\mu_\nu,
\end{equation}
where $H_{\mathbf{k}_\parallel}$ and $S_{\mathbf{k}_\parallel}$ are the Hamiltonian and the overlap matrix elements between Bloch basis functions, respectively.

%%DSP
In our case, the relevant initial wavepacket $\Phi(\mathbf{r})$  is one of the MOs of 
a single adsorbate, as it occurs for a localized excitation 
%%can be assumed when the system is probed by core-level spectroscopy
%%,\cite{10.1039/b719546j} so the initial state can be taken as the relevant orbital, $\Phi$, of %%the isolated molecule.
from a core level like in
core-hole-clock experiments.\cite{10.1039/b719546j} 
To avoid ambiguities, and to ensure their ``molecular character'', here
we use the MOs  
of an isolated, 
free-stading molecule (keeping the adsorbed geometry) 
as our initial wavepackets. Thus, 
$\Phi(\mathbf{r})=\sum_\mu b^\mu\phi_\mu(\mathbf{r}-\mathbf{R}_\mu)$, where
the $b^\mu$ are the coefficients of the MO in the atomic orbitals belonging to that molecule.
%the wavefunctions of the initial wavepackets
%% DSP (assuming that the orbital of interest is fully included in a single unit cell. 
The projection of the Green's function hence reads
\begin{eqnarray}\label{GPhiPhi}
G_{\Phi\Phi}(E)
%%&=& \int d\mathbf{r} d\mathbf{r}' \Phi^*(\mathbf{r}) G(\mathbf{r},\mathbf{r}',E) \Phi(\mathbf{r}')\nonumber\\
&=& \frac{1}{N_{\mathbf{k}_\parallel}} \sum_{\mathbf{k}_\parallel} \sum_{\lambda\mu\nu\rho} b_{\mathbf{k}_\parallel}^{\lambda*} S_{\lambda\mu\mathbf{k}_\parallel} G^{\mu\nu}_{\mathbf{k}_\parallel}(E) S_{\nu\rho\mathbf{k}_\parallel}b_{\mathbf{k}_\parallel}^{\rho},
\end{eqnarray}
where the   $\mathbf{k}_\parallel$ dependence of the coefficients 
$b_{\mathbf{k}_\parallel}^{\mu}= b^\mu \mathrm{e}^{ i\mathbf{k}_\parallel \cdot \mathbf{R}_\mu}$ is related to the phase conventions established in Eq.~(\ref{Blochbasis}).\cite{10.1088/0953-8984/14/11/302}
%%DSP unnecessary complications: 
%due to the coefficients of the Bloch eigenstate, $\Phi_{\mathbf{k}_\parallel}$, of the molecular Hamiltonian, computed within the same periodically repeated unit cell.
%%DSP your formula is basically based on a Wannier transformation. 
We notice here that
Eq.~(\ref{GPhiPhi}) is equivalent to a Wannier-like transformation of the 
MOs in our periodic molecular overlayer  $\Phi(\mathbf{r})=\frac{1}{N_{\mathbf{k}_\parallel}} \sum_{\mathbf{k}_\parallel} \Phi_{\mathbf{k}_\parallel}(\mathbf{r})$, producing the same result.
We observe that the width obtained from Eq.~(\ref{GPhiPhi}) as an
average of Green's functions at different $\mathbf{k}_\parallel$ cannot
be interpreted as the average of a $\mathbf{k}_\parallel$-dependent
linewidth. In particular,
%
%%exDSP
%In principle, one is tempted to interprete the width obtained from
%Eq.~(\ref{eq:GPhiPhi}) as an average over linewidth contributions
%coming from the interaction of the MO with substrate states with
%different $\mathbf{k}_\parallel$. However,
there is an additional contribution
in Eq.~(\ref{GPhiPhi})
due to the possible dispersion of the molecular bands within the overlayer, i.e., the energy 
of the MO-derived states might depend on $\mathbf{k}_\parallel$.
Notice that, in the experiment, the presence of a localized core excitation should exclude this possibility. However, the use of periodic boundary conditions in our calculations makes this point relevant even in the core-excited species. 
This additional width is not related
to charge transfer towards the substrate, but rather to hopping towards neighboring molecules, possibly mediated by the substrate.
This contribution should be avoided when focusing on isolated adsorbates, as it can be done by different schemes.\cite{10.1103/PhysRevB.76.235406}
Here, however, we have found that the interaction among neigboring molecules and the
corresponding molecular band width is negligible. Thus, it was not necessary 
to correct the  results obtained by Eq.~(\ref{GPhiPhi}).  It is also important
to stress that a 
sufficiently dense $\mathbf{k}_\parallel$ sampling, that scales inversely proportional to the lateral size of the supercell
used in the calculations,  is necessary to 
converge the results.
% 

%\subsection{Computational details}
While the formalism presented so far is rather general, we hereby specify the computational details for the simulations presented in the next Section.
We focus on isonicotinic acid (NC$_5$H$_4$COOH) adsorbed on the (110) surface of rutile TiO$_2$ (see Fig.\ref{fig:geometry}). According to the literature\cite{10.1016/S0009-2614(97)00070-5, 10.1016/S0009-2614(97)01143-3, 10.1016/S0039-6028(98)00278-7, 10.1146/annurev.physchem.58.052306.144054} we take a dehydrogenated, bidentate structure
%%DSP
with the O atoms of the molecule attached to the fivefold coordinated
Ti atoms of the surface, and the aromatic ring perpendicular to the substrate. A $(3\times1)$ surface periodicity is assumed. The molecular coordinates and those of the topmost two TiO$_2$ trilayers were determined by relaxing a 5-trilayer slab with molecules on both sides.
 The junction setup is constructed by adding four TiO$_2$ trilayers in bulk 
positions
 %%DSP
to the slab, obtaining the simulation cell shown in Fig.\ref{fig:geometry}. The left/right subsystems are separated by a vacuum region of $10$~{\AA}.
We have run DFT calculations with the Perdew-Burke-Ernzerhof\cite{10.1103/PhysRevLett.77.3865} approximation for exchange-correlation. A $(3\times2)$ $\mathbf{k}_\parallel$ mesh was adequate to determine the structural properties, while a denser $(12\times8)$ was adopted 
%%DSPin
to compute Eq.~(\ref{GPhiPhi}). Structural 
relaxations were performed with SIESTA using a double-$\zeta$ polarized (DZP) basis 
set.\cite{10.1088/0953-8984/14/11/302}
Several tests indicated that the 
TranSIESTA\cite{10.1103/PhysRevB.65.165401} results for the DOS of the system
were almost identical 
using  a DZP and a smaller double-$\zeta$ (DZ) basis set for the substrate. 
Therefore, we finally used the DZ basis for the Green's function calculations shown below.
%packages were run with DZP basis sets. Eventually, we checked that equivalent DOS could be %obtained by using the smaller DZ basis set for the substrate, so that was taken for the Green's %function calculations. 
The expansion coefficients for the MOs %%$b_{\mathbf{k}_\parallel}^{\mu}$ 
$b^{\mu}$
were extracted by a molecule-in-a-box calculation for the protonated species.
We used a protonated molecule to preserve its closed character,
although this hydrogen atom is lost upon adsorption. 
Fortunately, for the MOs that we considered here the contribution of this additional
hydrogen atom is negligible and the corresponding coefficients ($b^{\mu}$ with 
$\mu\in \text{extra-H}$) 
can be safely neglected without significantly modifying the distribution, shape and normalization
of the MOs.
We evaluate the DOS projected on the molecular orbital, $\rho_\Phi(E)$, from the imaginary part of the Green's function determined at complex energy $E+i\gamma/2$, where $\gamma$ adds to the total FWHM. So we fit the result with a lorentzian function with FWHM $\Gamma+\gamma$:
\begin{equation}
\rho_\Phi(E)\propto\frac{1}{\pi}\frac{(\Gamma+\gamma)/2}{
(E-E_\Phi)^2+[(\Gamma+\gamma)/2]^2},
\end{equation}
with $E_\Phi$ and $\Gamma$ as fitting parameters. We checked the independence of the results on $\gamma$ for various cases and chose the value $\gamma=20$~meV.

%\subsection{Technicalities}
Some additional technical aspects are presented next, highlighting the differences
with standard transport calculations using metallic leads.  In principle, the two left/right parts
of the ``contact'' in Fig.~\ref{fig:geometry} could be different, and especially one of them could be an undecorated surface, or an empty volume. However, in the
current version of SIESTA/TranSIESTA a preliminary slab calculation  is needed to get the Hamiltonian matrix elements so that, 
%%in the simplest approach, 
the left and right part 
%%DSP could be
are the upper and lower portions of such a slab (with unit cell as indicated by the solid line
in Fig.~\ref{fig:geometry}). Consequently, a symmetric configuration avoids spurious electric fields across the system.
Another issue concerns the connection of the electronic potentials of the contact region with that of the bulk electrodes.  The Fermi level can be chosen as a reference in a metallic system, but not in the case of a semiconductor like the one of interest here.  We instead align the potential (here the planar averaged one) deep inside the slab with that of the bulk one.\cite{10.1103/PhysRevB.67.155327} Differently, one could also take a deep level of TiO$_2$ as an energy reference. No external bias is applied in the present calculations.
As SIESTA uses pseudopotentials to describe core electrons, pseudopotentials for core-excited atoms\cite{10.1103/PhysRevLett.75.3166} were generated to describe the system in presence of a N 1s$^*$ atom, as it occurs in the XAS experiments that we want to address.\cite{10.1063/1.1586692}

\section{\label{sec:results}Results}
\begin{figure} \includegraphics[width=8.5cm]{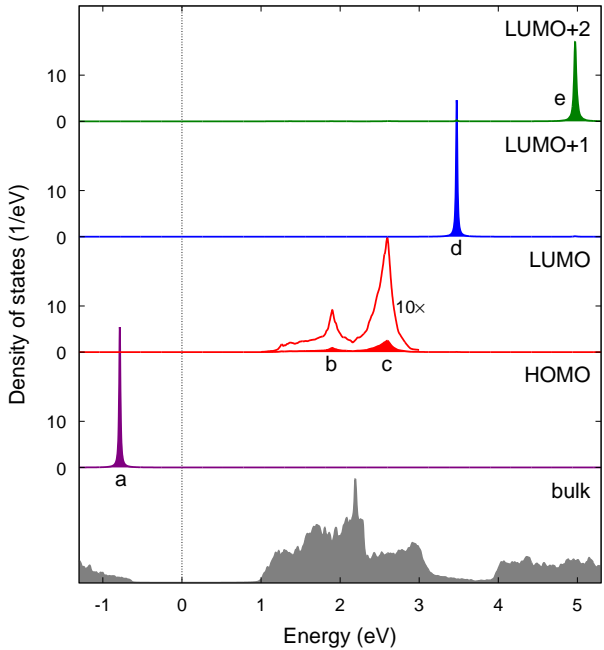}
\caption{\label{fig:PDOSGS}DOS for isonicotinic acid adsorbed on semi-infinite TiO$_2$(110), projected on the gas-phase molecular orbitals. The shaded area is the bulk TiO$_2$ DOS.}\end{figure}

\begin{figure} \includegraphics[width=8.5cm]{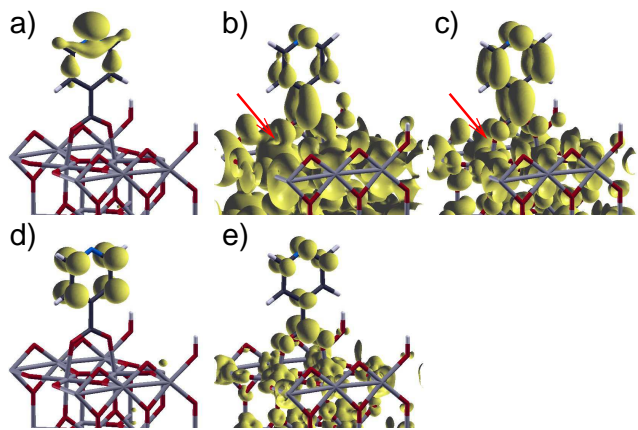}
\caption{\label{fig:LDOSGS}Local DOS corresponding to the (a) HOMO, (b-c) LUMO, (d) LUMO+1, and (e) LUMO+2 peaks in the DOS plot shown in Fig.~\ref{fig:PDOSGS}. 
%%DSP We should describe what the arrow is
The arrows in panels (b) and (c) highlight the absence and the existence, respectively,
of a nodal plane perpendicular to 
the Ti--O bonds, indicating the bonding/antibonding character of the 
substrate-molecule interaction.
}
\end{figure}

%\subsection{Ground state}
%%DSP
We start by discussing the properties of the system in the ground state, 
i.e., we do not introduce any explicit excitation in the system, but study the widths of the MOs
in the
KS spectrum corresponding to the ground state of the molecule/substrate system. 
The DOS projected on the gas-phase MOs ranging from the HOMO to the LUMO+2\footnote{As a matter of fact, the LUMO+2 of the adsorbed system corresponds to the LUMO+3 of the gas phase molecule, since the gas phase LUMO+2 
%%DSP
is localized at the H, eventually removed upon adsorption.} is displayed in Fig.~\ref{fig:PDOSGS}. The DOS of bulk TiO$_2$ is also shown, with the bandgap underestimated with respect to the experimental value of $3.0$~eV,\cite{10.1103/PhysRevLett.17.857} as expected 
%%DSP
using KS-DFT with standard semilocal functionals.
 From lower to higher energy, we can identify five features, named (a)-(e) in the figure, which are better discussed in conjunction with their real-space contribution (local DOS), integrated in small energy window ($\pm0.1$~eV) around the main peak. Those are reported in Fig.~\ref{fig:LDOSGS}(a)-(e), respectively.
The HOMO (a) close to the valence band edge shows no hybridization with substrate states and its
%%DSP wavefunction
density distribution is apparently the same as in the gas phase, with no weight on the anchoring -COO- group. For this state, which could be of potential interest for hole injection, we obtain $\Gamma=0$.
Conversely, the LUMO lies well within the conduction band. 
%%DSPHere, our method shows a level of detail which is unprecedented within this context.
Here, our method shows a level of detail beyond many approaches previously developed to  
determine resonant charge-transfer times.\cite{10.1021/jp2026847}
%%DSP 
For the LUMO we identify a behavior which has been well 
characterized for small adsorbates with strong electronic coupling
to the substrate:\cite{9780120078455} 
%%DSP
The MO splits in two components, (b) and (c), the former one with dominant
adsorbate-substrate bonding character and the latter one mostly antibonding (as the analysis of nodal structure shows, see Fig.~\ref{fig:PDOSGS}).  
Both components broaden in energy becoming resonances which extend spatially over the substrate. By fitting each peak independently, we obtain $\Gamma^\text{(b)}=80$~meV and $\Gamma^\text{(c)}=186$~meV, hence
%%DSP
 $\tau^\text{(b)}=8$~fs for (b) and $\tau^\text{(c)}=4$~fs for the most intense component (c).
These values can be compared to those given by Martsinovich and Troisi.\cite{10.1021/jp2026847} They adopted a Hamiltonian-partitioning technique to compute the 
coupling elements between the molecule and substrate, and from that to obtain the lifetime for the LUMO state (taken as a whole). Their value, $0.68\pm0.41$~fs, corresponds to $\Gamma\approx1$~eV which is consistent with
%% DSP: the overall extension seems larger, however the splitting... 
%% the overall extension in energy of the LUMO state that we report in \ref{fig:PDOSGS}. 
the splitting between the LUMO components that we report in Fig.~\ref{fig:PDOSGS}.
%%DSP
Therefore, our calculation confirms a
strong  interaction between the 
isonicotinic acid and the TiO$_2$ substrate 
as that reported by Martsinovich and Troisi. However, our simulations also point out
that such interaction cannot be  interpreted solely in terms of a single resonant lifetime, since it determines
a splitting of the LUMO
to two components which can be pumped independently.
Next, the LUMO+1 (d), 
%%DSP
which falls in a 
region with a small TiO$_2$ DOS and has no weight on the -COO- termination of the molecule, shows no hybridization with the substrate, with $\Gamma=0$. The LUMO+2 couples to substrate states, and a single peak can be identified in this case for which we obtain $\Gamma=15$~meV ($\tau=44$~fs).
Finally, from Fig.~\ref{fig:PDOSGS} one can also notice that the adsorption does not introduce a rehybridization among molecular states,
%%DSP
 which seem well described in terms of those of the gas phase molecule.
All the values of the lifetimes for the ground state molecule
are collected in Fig.~\ref{tab:GammaTau}. They could refer to the lifetime of electrons either added to the system or promoted to unoccupied states
%%DSP
from extended
MOs, where the role of the electron-hole interaction is less determinant and often neglected.

\begin{table}
\begin{tabular}{lccccc}											
Ground state	&	HOMO	&	LUMO (b)	&	LUMO (c)	&	LUMO+1	&	LUMO+2	\\
$\Gamma$ (meV)	&$	0	$&$	80	$&$	186	$&$	0	$&$	15	$\\
$\tau$ (fs)	&		&$	8	$&$	4	$&		&$	44	$\\
\hline											
N 1s$\rightarrow$valence	&		&		&		&	LUMO+1	&	LUMO+2	\\
$\Gamma$ (meV)	&		&		&		&$	7	$&$	15	$\\
$\tau$ (fs)	&		&		&		&$	93	$&$	44	$
\end{tabular}											
\caption{\label{tab:GammaTau}Linewidths (FWHM, $\Gamma$) and transfer time, $\tau$, for molecular orbitals of isonicotinic acid on TiO$_2$(110) as in a neutral and core-excited configuration.}\end{table}

Before considering the influence of a core-level excitation, it is worth stressing that relatively
small widths, 
%DSP LUMO+1 has strictly zero... 
like that of the LUMO+2 state reported here, are difficult to determine from the DOS of molecules adsorbed on a slab or cluster. In such a system the states (artificially quantized in the direction perpendicular to the surface) should be dense enough to allow for working out a lorentzian profile. 
Consider as an example the energy window just above the conduction band edge, where the bulk DOS is about $3$~states/eV/spin/unit cell (6 atoms). Hence with a $(3\times1)$ surface periodicity we need more than $50$ TiO$_2$ trilayers 
to achieve an average distance between electronic levels of $2$~meV, 
%%DSP for the substrate DOS to be denser than ($2$~meV)$^{-1}$ 
which allows obtaining five states (at each $\mathbf{k}_\parallel$) within the FWHM of a resonance with $\Gamma=10$~meV. Thicker slabs would then be necessary for even narrower resonances.
% DOS vs num of tril: (TiO2)2	(48 electrons)	(6 states in a 2eV window)	3states/eV/UC
% 3x1 => 9states/eV/trilayer;	0.002 eV/state => 500 states/eV => >50 trilayers

\begin{figure} \includegraphics[width=8.5cm]{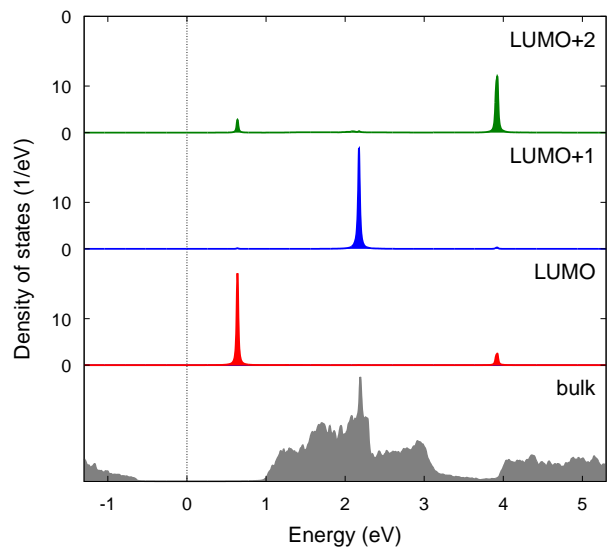}
\caption{\label{fig:PDOSCH}DOS for isonicotinic acid adsorbed on semi-infinite TiO$_2$(110), projected on the gas-phase molecular orbitals, with a N 1s electron excited to valence. The shaded area is the bulk TiO$_2$ DOS. Compare with Fig.~\ref{fig:PDOSGS}.}\end{figure}

\begin{figure} \includegraphics[width=8.5cm]{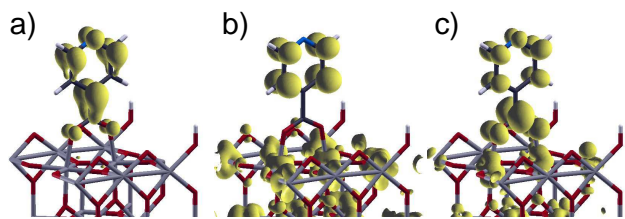}
\caption{\label{fig:LDOSCH}Local DOS corresponding to the (a) LUMO, (b) LUMO+1, and (c) LUMO+2 peaks in the DOS plot shown in Fig.~\ref{fig:PDOSCH}, with N 1s electron excited to valence.}\end{figure}

%\subsection{With core hole}
%%DSP
We now turn to the case of a molecule excited by X-ray radiation. We considered excitations
from the N 1s state, corresponding to a binding energy of  $\sim400$~eV, as occurs in experimental measurements by the core-hole-clock method.\cite{10.1063/1.1586692} 
%%DSP
In the simulation, this is obtained by promoting a N 1s electron from the core to a
valence state. 
%%DSP
The core-hole is included in the pseudopotential that now accounts for the interaction of 
valence electrons with a core-excited N 1s$^*$ ion. 
The excited electron is explicitly included in the self-consistent calculation (so that the system
remains neutral) populating the molecular LUMO and, due to the
presence of the N core-hole, remains strictly 
localized within the molecule (see below). The additional
%%DSP
positive nuclear charge lowers the effective potential in the molecular region, shifting the molecular orbitals to lower energies. See Fig.~\ref{fig:PDOSCH} where the 
corresponding projected DOS is shown.
In close agreement with the experiments\cite{10.1063/1.1586692} the LUMO is brought down 
%%DSP
inside
the energy gap, so that no resonant charge transfer can take place from this state
of the molecule with a core-excited N atom. The LUMO's density 
distribution is shown in Fig.~\ref{fig:LDOSCH}(a), in which the coupling with substrate states is absent,
%%DSP
in clear contrast with the ground-state molecule. 
As a consequence of the core-hole attraction, the LUMO is now polarized towards 
the N atom, 
as can be observed by comparing Fig.~\ref{fig:LDOSCH}(a) with Fig.~\ref{fig:LDOSGS}(b)-(c). 
Indeed, with respect to the MOs of the free molecule, we can appreciate some
rehybridization of the states: see the LUMO and LUMO+2 peaks in Fig.~\ref{fig:PDOSCH}.

At variance with the neutral-core case shown in Fig.~\ref{fig:PDOSGS}, the LUMO+1 and LUMO+2 are also significantly downshifted. In particular, the LUMO+1 enters a region of large density of substrate states with which to couple, as can be seen in Fig.~\ref{fig:LDOSCH}(b). We obtain a width $\Gamma=7$~meV ($\tau=93$~fs). Notice how the symmetry and 
%%DSP
spatial distribution of the orbital strongly influence the transfer rate. In fact the LUMO+1
 of the core-excited case (Fig.~\ref{fig:LDOSCH})
%%DSP
 lies in the same energy region as the LUMO of the ground state system shown in Fig.~\ref{fig:PDOSGS}, but the resonant lifetime is at least one order of magnitude larger. 
For the LUMO+2, whose LDOS is depicted in Fig.~\ref{fig:LDOSCH}(c), one obtains coincidentally the same width as that for the neutral-core molecule ($\tau=44$~fs), although the orbital couples to different substrate states.
We mention that the HOMO, not reported here, lies at much lower energies, because of its large weight around the N atom (recall Fig.~\ref{fig:LDOSGS}(a)) where 
the 
%%DSP
attraction by the
%%DSP
core-hole is more effective.


\section{\label{sec:discussion}Discussion}
%\subsection{Need of e-h interaction}
Our results, and in particular the comparison between Fig.~\ref{fig:PDOSGS} and Fig.~\ref{fig:PDOSCH}, highlights the importance of excitonic effects in the calculation of the resonant lifetime, in comparing to measurements 
%%DSP
obtained
%%DSP
with the core-hole-clock method.
Excitonic effects are also important for the transfer of electrons excited by visible light, where the simplifications adopted here 
%%DSP for the hole wavefunction 
due to the localized character of the core-hole
do not hold. Other methods, like time-dependent DFT or the two-particle Bethe-Salpeter equation could be used instead.\cite{10.1103/RevModPhys.74.601} 
%%DSP
However, the large computational cost of such approaches limits their applicability to the evaluation of optical spectra for relatively small systems,\cite{10.1088/0957-4484/19/42/424002} while no implementation for semi-infinite substrates (to access transfer times) exists to the best our knowledge. Hence electron-hole interaction is most often neglected in computing resonant linewidths, as we do here for the neutral-core case.

%\subsection{Disagreement with experiments?}
Owing to the large excitonic effect 
%%DSP
induced by the presence of the N core-hole, the agreement of the linewidth of isonicotinic acid with the experimental one\cite{10.1063/1.1586692} claimed in Ref.~\cite{10.1021/jp2026847}, where the  ground-state configuration was used, is fortuitous.
%%DSP
Our analysis instead correctly suppresses resonant transfer from the LUMO state, and comparison to the LUMO+1 and LUMO+2 states should be done instead. For these states (especially the LUMO+2 with largest signal/noise ratio) an upper limit of $5$~fs was measured.\cite{10.1063/1.1586692} Since we are focusing on the resonant part of the injection process, our estimates are necessarily upper bounds of the complete (theoretical) transfer time, 
%%DSP
that will be decreased due to additional
inelastic contributions (due, e.g., to electron-phonon and electron-electron interactions\cite{10.1103/PhysRevB.80.195419}). Still, our calculated resonant charge transfer time is
one order of magnitude larger than that in experiments, a discrepancy 
which
% is difficult to ascribe solely to inelastic processes. 
might be due to other effects not included in our estimate, as we further discuss.
%%{\bf [DSP]} % Daniel this is a point where any work on resonant transfer could be criticized, would you change anything?
%%DSP I think the discussion is fine as it is now. 

\begin{figure} \includegraphics[width=8.5cm]{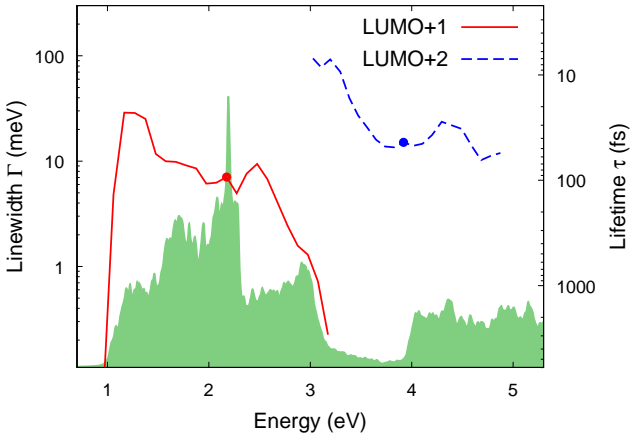}
\caption{\label{fig:shiftH}Linewidth and lifetime of the molecular orbitals of isonicotinic acid with one N 1s electron excited into the LUMO, as a function of the energy of the molecular orbitals. Bullets mark the self-consistent values, and the lines are the results obtained using a post-self-consistent rigid shift of the molecular Hamiltonian operated by Eq.~(\ref{eqshiftH}). The shaded area is the bulk TiO$_2$ DOS.}\end{figure}

%\subsection{DFT levels}
In common to most other studies in the related literature (an exception was recently given\cite{10.1063/1.4809994}) our approach takes single-particle energy levels from KS eigenvalues, while DFT is a ground state theory. Luckily enough, some errors 
%%DSP
in the 
KS-DFT spectrum tend to cancel out. This occurs for the TiO$_2$ band gap and the HOMO-LUMO gap in the molecule, both being underestimated. 
Due to a fortuitous coincidence, the relative position of molecular and substrate states shown in Fig.~\ref{fig:PDOSCH} seems to be in rather good agreement with the one reported in the experiments in Ref.~\cite{10.1063/1.1586692}. However, since {\em a priori} there is no guarantee that proper energy alignment stems from error cancellations, it is important to understand how the results depend on such alignment.
%\subsection{Shift of energy levels}
For this reason, we rigidly shift the molecular orbital energies with respect to the substrate bands
%% by means of a scissor-like operator acting
as a  post-self-consistency correction. 
%%on the adsorbate only. 
%%DSP Yes, I would not call this "scissor". 
%%%{\bf [ALL]} % The scissor operator is between occupied and unoccupied states. Here we don't really bother about occupied ones... shall we call it otherwise (simply “operator”)?
This is implemented by adding the energy shift $\Delta\epsilon$ to the terms of the Hamiltonian matrix belonging to the adsorbate:
\begin{eqnarray} \label{eqshiftH}
H_{\mu\nu\mathbf{k}_\parallel}=H_{\mu\nu\mathbf{k}_\parallel}
+ S_{\mu\nu\mathbf{k}_\parallel} \Delta \epsilon,
\mu,\nu \in\text{molecule}. 
\end{eqnarray}
Next we compute the electronic structure with values of $|\Delta\epsilon|$ up to $\approx1$~eV and the corresponding lifetime of LUMO+1 and LUMO+2 states (in presence of the N 1s excitation). The results are reported in Fig.~\ref{fig:shiftH} as a function of the energy of the MOs, 
showing a moderate dependence on energy, except for the LUMO+1 at about $3$~eV.
%%DSP
According to these data, the misalignment of the MOs with respect
to the TiO$_2$ bands does not appear to be at the origin of the large discrepancy between the 
measured and the calculated charge transfer times in this system.
It is also interesting to note that, although the lifetimes in Fig.~\ref{fig:shiftH} approximately
follow the profile of the TiO$_2$ substrate DOS, the symmetry and spatial 
distribution of both, the molecule and substrate states, plays a key role in 
determining the values of the resonant charge transfer times.  

%\subsection{Neglecting ionic motion}
Finally, we consider the effects of neglecting a finite temperature 
and those related to the use of ground state geometry. 
%%DSP this is misleading. You can frozen ions always if the electron transfer is fast,
%% you refer to a different thing.
%we recall that a description of resonant electron transfer at frozen atomic coordinates is properly descriptive of the low-temperature limit. 
As temperature is increased, we should consider different effects induced by the dynamics of the system.
First, the core-level excitation could occur for molecules in different structural configurations than the energy minimum. This could open up new decay channels which might vanish in the high-symmetry structure depicted in Fig.~\ref{fig:geometry}. This feature was studied extensively in various contexts, by making averages over snapshots of molecular dynamics simulations.\cite{10.1021/jz200191u}
A detailed analysis in this direction goes beyond the realm of the present work. However, as an example, we computed the effect of a rigid $30^\circ$ 
%%DSProtation 
tilt of the molecule around the [001] axis passing through the O atoms of the -COO group (which is compatible with a polar angle of $0\pm40^\circ$ observed experimentally\cite{10.1016/S0039-6028(03)00827-6}): the coupling of the $\pi$ electron system to the substrate is 
enhanced, resulting in a significant reduction of the LUMO+2 lifetime to 
about $1/3$ of the calculated value for the molecule remains perpendicular to the surface. 
Therefore, the presence of modified geometries in which the
molecule interacts more strongly with the TiO$_2$ substrate, 
might be one
of the reasons behind the small charge-transfer times measured for isonicotinic acid,\cite{10.1063/1.1586692} and a way to reconcile 
the experiment with our theoretical estimations.

\section{\label{sec:conclusions}Conclusions}
The resonant lifetimes of the 
%%DSP electron 
molecular states of isonicotinic molecules adsorbed 
%%DSP at
on the surface of rutile TiO$_2$(110) were evaluated from first principles.
Using a Green's function based methodology we could take 
into account the continuum density of states of the semi-infinite substrate.
This is not accessible by slab or cluster models and provides 
%DSP increased 
a very detailed picture of the hybridization between molecule and substrate states.
%A Green's function based procedure to describe the semi-infinite host electronic structure was %presented. 
%%DSP
%%DSP 
For an isonicotinic acid molecule in its ground state, 
the Kohn-Sham LUMO resonance broadens and splits into two components, 
with linewidths corresponding to short lifetimes ($4$ and $8$~fs, respectively).
However, in the presence of a core-excited atom (as occurs in measurements of electron transfer by core-level spectroscopy),
%%DSP, and here introduced in the calculation), 
electron injection from the LUMO is suppressed as it enters the 
%%DSPsemiconductor 
substrate band gap. 
%%%
%%%Hence, the apparent agreement between 
%%%calculations of the resonant charge-transfer rates from the LUMO to the TiO$_2$ substrate
%%%ignoring the effect of core-excited atoms,\cite{10.1021/jp2026847} and those measured using core-hole-spectroscopy,\cite{10.1038/nature00952, 10.1063/1.1586692} 
%experimental results, sometimes claimed in the literature, is only 
%%%is purely fortuitious.
%%%
Resonant transfer is still possible for higher lying states, for which longer times ($93$~fs and $44$~fs for the LUMO+1 and LUMO+2, respectively) are calculated for the most stable adsorption configuration. Such values are at least one order of magnitude larger than the 
lifetimes of a few femtoseconds found in the experiment, which also comprise inelastic contributions (not taken into account in the present work).
%%DSP Although the resonant channel is not the only possible electron transfer mechanism. 
The dependence on the relative alignment of molecular and substrate states was investigated and demonstrated to have a moderate effect on the results reported here. 
%Although, the resonant channel is not the only possible electron transfer mechanism, it seems
%difficult to attribute the origin of this large discrepancy solely to additional 
%inelastic contributions. For this reason we
We speculate that other adsorption configurations with an increased molecule/substrate interaction, different from the relaxed ground-state geometry  used here,  could
be abundant in the experimental situation. 

Our work
implements Green's function methods to the calculation of the lifetime of molecular states at an extended substrate.
It highlights the importance of incorporating the
excitonic effects associated  with the presence of core-excited species
when comparison with experimental information obtained using 
the core-hole-clock method is pursued.
It further stresses that various types of experiments might provide contrasting
charge transfer times since, in some cases, they look into
very different spatial distributions and couplings to the substrate of the involved MOs.

% End of the paper
\acknowledgement %%%JPCC
%%% \section*{ACKNOWLEDGMENTS} %%%APS
We acknowledge support from the MIUR of Italy through PRIN project DSSCX (n. 20104XET32). Computational resources were made available in part by CINECA (application code HP10C0TP0R).
CM thanks CARIPLO Foundation for its support within the PCAM European Doctoral Programme and Pirelli Corimav for his PhD scholarship.
DSP acknowledges support the Basque Departamento de Educaci\'on, UPV/EHU (Grant No. IT-366-07), 
the Spanish Ministerio de Ciencia e Innovaci\'on (Grant No. FIS2010-19609-C02-02), the ETORTEK program funded by the Basque Departamento de Industria and the
Diputaci\'on Foral de Guipuzcoa, the EU through the FP7 PAMS project and the German DFG through SFB 1083.

\bibliography{lifetime}
\end{document}